\documentclass[reprint,
superscriptaddress,
amsmath,
amssymb,
aps,
prb,
nobibnotes
]{revtex4-2}

\usepackage{xurl}
\usepackage{lipsum}
\usepackage{graphicx}
\usepackage{dcolumn}
\usepackage{bm}
\usepackage{comment}
\usepackage{upgreek}
\usepackage[usenames,dvipsnames,table]{xcolor}
\usepackage[mathlines]{lineno}
\usepackage{multirow}
\usepackage{siunitx}
\usepackage[colorlinks]{hyperref}
\usepackage{amssymb,amsmath,amsthm,enumitem}
\usepackage{soul}
\usepackage{nameref}
\usepackage{chngcntr}
\usepackage{makecell}

\DeclareUnicodeCharacter{2212}{-}
\DeclareGraphicsExtensions{.pdf,.png,.jpg}

\definecolor{bluegray}{RGB}{40,180,160}
\definecolor{navygray}{RGB}{110,140,170}
\definecolor{meadowgreen}{RGB}{0,128,0}
\definecolor{magenta}{RGB}{255,0,255}
\definecolor{lightgrey}{RGB}{200,200,200}
\definecolor{C0}{RGB}{31,119,180}
\definecolor{C1}{RGB}{255,127,14}
\definecolor{C2}{RGB}{44,160,44}
\definecolor{R5}{RGB}{31,119,191}
\definecolor{R4}{RGB}{255,152,28}
\definecolor{R3}{RGB}{44,160,44}
\definecolor{R2}{RGB}{214,39,40}
\definecolor{R1}{RGB}{148,103,189}
\definecolor{R0}{RGB}{140,86,75}

\hypersetup{
citecolor={bluegray}, 
linkcolor={navygray}, 
}

\DeclareSIUnit[]\samples
{S}

\begin{document}

\title{The spatial correlation of radiation-induced errors \\ in superconducting devices decays over a millimeter}

\author{Francesco~Valenti}
\thanks{First two authors contributed equally.}
\affiliation{IQMT,~Karlsruhe~Institute~of~Technology,~76131~Karlsruhe,~Germany}
\affiliation{Current address: IBM Quantum, IBM T. J. Watson Research Center, Yorktown Heights, 10598 NY, USA}

\author{Anil~Murani}
\thanks{First two authors contributed equally.}
\affiliation{IQMT,~Karlsruhe~Institute~of~Technology,~76131~Karlsruhe,~Germany}
\affiliation{Current address: Alice \& Bob, 75015 Paris, France}

\author{Patrick~Paluch}
\affiliation{IQMT,~Karlsruhe~Institute~of~Technology,~76131~Karlsruhe,~Germany}
\affiliation{PHI,~Karlsruhe~Institute~of~Technology,~76131~Karlsruhe,~Germany}

\author{Robert~Gartmann}
\affiliation{IPE,~Karlsruhe~Institute~of~Technology,~76131~Karlsruhe,~Germany}

\author{Lukas~Scheller}
\affiliation{IPE,~Karlsruhe~Institute~of~Technology,~76131~Karlsruhe,~Germany}

\author{Richard~Gebauer}
\affiliation{IPE,~Karlsruhe~Institute~of~Technology,~76131~Karlsruhe,~Germany}

\author{Robert~Kruk}
\affiliation{INT,~Karlsruhe~Institute~of~Technology,~76131~Karlsruhe,~Germany}

\author{Thomas~Reisinger}
\affiliation{IQMT,~Karlsruhe~Institute~of~Technology,~76131~Karlsruhe,~Germany}

\author{Luis~Ardila-Perez}
\affiliation{IPE,~Karlsruhe~Institute~of~Technology,~76131~Karlsruhe,~Germany}

\author{Ioan~M.~Pop}
\email{ioan.pop@kit.edu}
\affiliation{IQMT,~Karlsruhe~Institute~of~Technology,~76131~Karlsruhe,~Germany}
\affiliation{PHI,~Karlsruhe~Institute~of~Technology,~76131~Karlsruhe,~Germany}
\affiliation{Physics~Institute~1,~Stuttgart~University,~70569~Stuttgart,~Germany}

\begin{abstract}
We perform nanosecond-resolution multiplexed readout on six same-chip superconducting microwave resonators. This allows us to pinpoint the position of ionizing radiation impacts on the chip by measuring the differential time of flight of the generated phonons, inducing correlated errors in the device, thereby implementing an on-chip seismic array. We correlate the phase response of each resonator | a proxy for the absorbed energy | to the distance from the impact point to uncover a millimetric decay length for the phonon-mediated radiation poisoning.
\end{abstract}

\maketitle

\section{Introduction}

The quest towards fault-tolerant superconducting quantum computers has recently branched into a multidisciplinary effort between the superconducting quantum electronics and the high-energy physics communities. The reason is that ionizing radiation has been found to have detrimental consequences for quantum applications, as it can lead to reshuffling of the microscopic environment \cite{thorbeck_radiation} and phonon-mediated quasiparticle poisoning of the superconducting film \cite{Vepslinen2020impact,valenti_cardani_demetra,larson2025}. The latter has been shown to result in correlated errors in both resonators \cite{Swenson2010,Karatsu2019} and qubits \cite{Wilen2021,McEwen2022,yelton_modeling},  undermining the hypothesis of non-correlated errors for quantum error correction. Despite being partially mitigated by superconducting gap engineering \cite{HenriquesValenti2019,martinis2020,McEwen2024Feb}, this quasiparticle poisoning poses a challenge to hardware frugal architectures. In particular, it has recently been observed that the logical error per cycle in a quantum processor was limited by high-energy impacts in a 1D repetition code \cite{McEwen2022}, and correlated chip-level errors are still present in the latest error correction demonstration \cite{Acharya2024Aug}. Understanding radiation-induced, phonon-mediated quasiparticle poisoning in superconducting quantum circuits constitutes an outstanding challenge in the path towards fault-tolerant quantum processors.

In this work, we employ nanosecond-resolution, in-house-developed electronics to perform multiplexed reflection measurements on six same-chip resonators. This allows us to use the arrival time differential for pair-breaking phonons to estimate the on-chip position of the ionizing impact, thereby implementing a superconducting seismic array, in the spirit of Refs.~\cite{Swenson2010,moshe_speed}. Furthermore, we resolve the fast rise of the resonator response and use it as a proxy for the energy deposited into the resonators. Equipped with these two quantities, we uncover a millimetric correlation length of correlated errors induced by quasiparticle bursts generated by ionizing impacts.

 \begin{figure*}[t!]
    \includegraphics[width=1\textwidth]{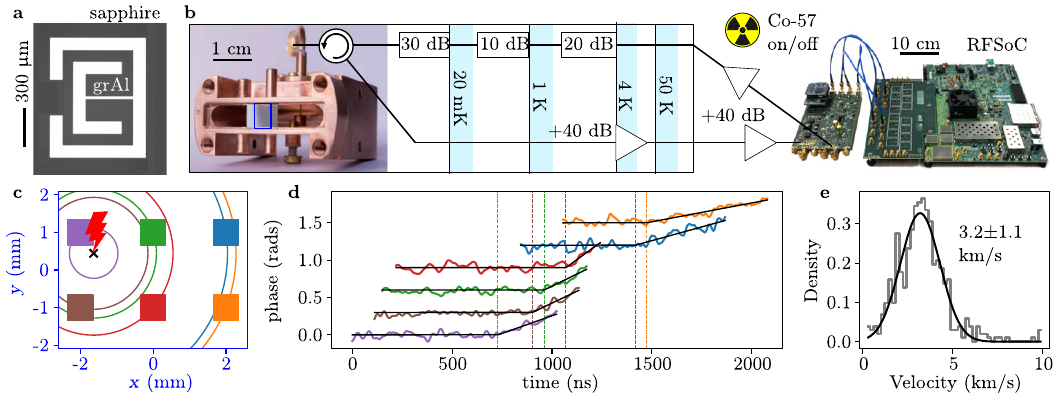}
    \caption{\textbf{Experimental setup.}  \textbf{a}, micrograph of one of the six granular aluminum (grAl, white) superconducting resonators on a sapphire (gray) chip. The lumped design, inherited from Ref.~\cite{winkel_transmon}, comprises an inductive wire, which acts as the phonon absorber, while the width of the electrode pitch can be swept to change the resonator frequency. \textbf{b}, schematics of cryostat and microwave wiring. The seismic array chip is glued into a copper waveguide mounted at the mixing chamber stage of a dilution cryostat. The signal is attenuated from room temperature at successive temperature stages, and recovered via low noise amplifiers at 4~K and at room temperature. The drive and readout electronics consist of a commercial Radio Frequency System-on-Chip (RFSoC) board with custom firmware for baseband signal generation, real-time frequency comb demodulation, and online data processing, along with a custom analog front-end board that converts between baseband and radio frequency using a superheterodyne mixer approach (cf. Appendix~\ref{app_rf} for details). A cobalt-57 source (trefoil symbol) can be put next to the cryostat to raise the radioactivity levels for control experiments. \textbf{c}, schematics of the superconducting seismic array, composed of a $3\times2$ grid of resonators (colored squares, roughly to scale | cf. panel a) spaced $2$~mm apart on a sapphire chip (blue rectangle in panel b). An ionizing impact (red bolt) generates a phonon wavefront that propagates isotropically from the impact point (black cross) via a quasi-diffusive process (see main text for details). The arrival of the wavefront at each resonator (color-coded circles) can be used to reconstruct the impact position, as well as to infer the wavefront propagation velocity. \textbf{d}, phase of the complex frequency response of each resonator/seismometer, used to reconstruct the impact position as indicated in panel \textbf{c}, and using the same color palette. Traces are offset in steps of 0.3 radians for clarity, increasing by order of arrival time (dashed vertical line). \textbf{e}, histogram of reconstructed propagation velocity of the phonon wavefront for all registered impacts reported in Fig.~\ref{fig_length}a in $0.1$~km/s bins (gray), together with fit to a Gaussian distribution (black).} \label{fig_setup}
\end{figure*}

 \begin{figure*}[t!]
    \includegraphics[width=1\textwidth]{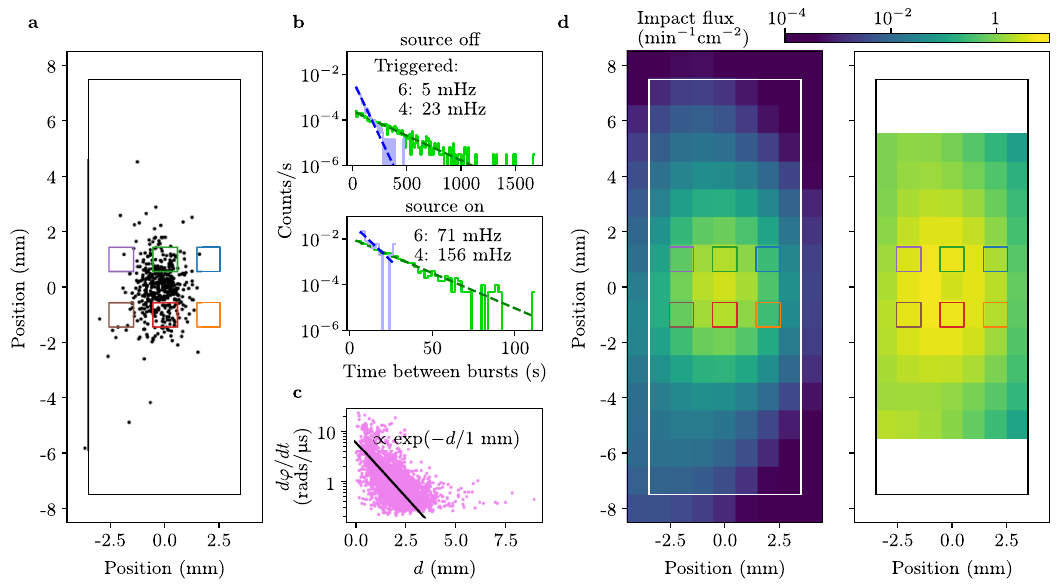}
    \caption{\textbf{Evidence of a correlation length}. \textbf{a}, scatter plot of reconstructed positions of impacts on the chip (black) over a $\sim$27 hour measurement window. The chip boundaries are shown as an empty rectangle; resonators (roughly to scale) are shown as empty squares (cf. color palette of Fig.~\ref{fig_setup}c,d). \textbf{b}, distributions of waiting times between successive bursts from environmental radiation (top) and with the addition of a radioactive source (bottom), with $10$ s and $2$ s binning, respectively. Impacts correlated across six (four) resonators are shown in green (blue) histograms. Exponential fits (solid lines) are used to extract the reported rates. All other panels in this figure report data measured without the source. \textbf{c}, slope of the phase rise as a proxy for the energy deposited in the resonators by the reconstructed impacts, as a function of their distance from each resonator (pink), together with an exponential decay with $1$ mm characteristic length (black). \textbf{d}, reconstructed flux of ionizing particles impinging on the chip, reported both as-is (left) and after applying a scaling with the decay length of panel c (cf. main text for discussion). The color bar is shared across the two panels.} \label{fig_length}
\end{figure*}

\section{Seismometry} \label{sec_seismo}

 High-energy particle impacts (``impacts" from now on) on the substrate generate phonons by e.g. generating electron/hole pairs that recombine, or by displacing nuclei. 
 Phonons impinging on a superconducting resonator with energy above the superconducting spectral gap break Cooper pairs and generate out-of-equilibrium quasiparticles (a so-called quasiparticle burst, and simply ``burst" from now on). Typically, following a burst, an energy in the eV | keV range is deposited in the superconducting device~\cite{valenti_cardani_demetra}. The resulting increase in kinetic inductance lowers the resonant frequency, and the resonator plays the role of a kinetic inductance detector~\cite{day2003broadband,Swenson2010}. The resonant frequency shift is related to the amount of energy coupled by phonons into the superconducting resonator 
 \begin{equation} \label{eq_shift}
    \delta E = \underbrace{- \frac{4}{\alpha} \frac{\delta f_0}{f_0}}_{= \delta x_\mathrm{QP}} n_\mathrm{CP} \Delta_0 V,
\end{equation}
where $\alpha$ is the kinetic inductance fraction, $\delta f_0 / f_0$ is the relative resonant frequency shift, $\delta x_\mathrm{QP}$ is the fractional shift in quasiparticle density, $n_\mathrm{CP}$ is the Cooper pair density, $\Delta_0$ is the superconducting gap, and $V$ is the mode volume. 

The energy of incoming radiation is in the keV|MeV range \cite{valenti_cardani_demetra,Cardani2023Jan}, orders of magnitude greater than the maximum phonon energy (e.g. circa $0.1$~eV for sapphire \cite{sapphiredos}). This results in frequent scattering and a so-called quasi-diffusive process for phonon propagation: a “ball” of phonons
expands isotropically at some fraction of the
speed of sound \cite{ogburn2008thesis}. This is the mechanism inducing correlated bursts and errors in superconducting devices on the same chip.

In the spirit of Ref~\cite{Swenson2010}, we use the time differential of wavefront arrival on each of the $N$ resonators which detect a burst to reconstruct the impact position. We use \texttt{scipy}'s least squares method \cite{scipylsq} to solve the system of $N$ equations
\begin{align} \label{eq_sys}
    (x-x_k)^2 + (y-y_k)^2 - (t_{1k}+t)^2c^2 = 0,\\
    \text{with } t = \sqrt{(x-x_1)^2+(y-y_1)^2}/c \nonumber,
\end{align}
where $x,y$ is the impact point, $c$ is the propagation velocity, and $x_k,y_k$ are the coordinates of the $k$-th resonator, where $k$ denotes the arrival order ($x_1,y_1$ is the position of the first resonator reached by the wavefront), and $t_{1k}$ is the differential time of arrival between the first and $k$-th resonator (cf. Appendix~\ref{app_position}). The minimum number of resonators to solve the system is $N=4$, given that we don't know the time of the original impact, and the propagation velocity is possibly a function of the unknown impact energy \cite{moshe_speed} |  for $N>4$ this is an overcomplete system. 

Once the phonon wavefront reaches the resonator, quasiparticles are generated on a timescale that is faster than all other relevant dynamics in the resonator. This is justified by the energy of the phonons being orders of magnitude larger than the superconducting spectral gap \cite{sapphiredos} and the electron-phonon interactions times being on the order of a nanosecond \cite{kaplan76}. Furthermore, the relaxation process is slower than all other relevant dynamics in the resonator, and can be as long as seconds~\cite{Grunhaupt2018QPgral}. This entails that we can treat the induced frequency shift as a step function in time. In the immediate aftermath of a burst, the complex microwave reflection or transmission coefficient of the resonator starts on a trajectory towards the new steady state corresponding to the shift. We resolve the initial phase slope following a burst and use it as a proxy for the energy deposited into the resonator (cf. Appendix \ref{app_reso}). As such, by concomitantly monitoring the microwave response of several resonators, we can resolve both the impact position and a scale of the wavefront energy at each detector site, thereby implementing an on-chip seismic array. 

\section{Experimental setup} \label{sec_setup} \label{sec_setup}

We fabricate a chip with $N=6$ superconducting resonators in a multiplexed readout scheme, each playing the role of a seismometer. For the superconducting film we employ granular aluminum (grAl), because we can tune its kinetic inductance during deposition by controlling the flow of oxygen \cite{Rotzinger_2016} while preserving $\mathcal{O}(10^5)$ internal quality factors in the single photon regime \cite{Grunhaupt2018QPgral} even for highly oxidized samples. Higher kinetic inductance results in detectors with better responsivity (cf. Eq.~\eqref{eq_shift}). The resonator design, inherited from Ref.~\cite{winkel_transmon} and shown in Fig.~\ref{fig_setup}a, consists of two electrode pads connected by a wire acting as the inductor. The pitch in the top electrode can be swept to change the resonant frequency, in the case of our devices spanning the $8.7-9$~GHz range. The $55$~nm film is patterned on top of $c$-plane $330\;\upmu$m thick sapphire with a single lift-off optical lithography step. We measure the room-temperature resistance using test stripes deposited on the wafer in the same evaporation step and use the Mattis-Bardeen formula \cite{Rotzinger_2016} to infer a kinetic inductance $L_K = 155$~pH$/\square$. By calculating the geometric inductance $L_G$ from design parameters \cite{valenti2019interplay} we estimate a kinetic inductance fraction $\alpha = L_K / (L_K + L_G) \approx 0.4$. 

A schematic of the experimental setup is shown in Fig.~\ref{fig_setup}b. The sapphire chip hosting the resonators is glued into a copper waveguide (same design as Ref.~\cite{winkel_transmon}) using a 1:1 mixture of epoxy resin and silver powder with a $4\;\upmu$m mean particle size. The suspended chip couples to the TE10 mode of the waveguide. The waveguide cutoff is at 4.5 GHz, well below the resonator frequencies. The waveguide is attached to the mixing chamber plate of a dilution cryostat. The microwave lines in the cryostat are attenuated for a nominal total of 60 dB on the way in. The signal is routed to and from the waveguide by means of a circulator, and amplified on the way up by both a High Electron Mobility Transistor (HEMT) at the 4~K plate and a room-temperature amplifier. We can artificially raise the environmental radioactivity by putting a cobalt-57 source in the proximity of the cryostat.

We developed in-house printed circuit boards for both the baseband fanout and the Radio Frequency (RF) mixing stage. Details on the electronics can be found in Appendix~\ref{app_rf} and Ref.~\cite{Gartmann_2022}. The Field Programmable Gate Array (FPGA) board generates a comb of six tones, with a total bandwidth of circa $300$~MHz. This is mixed up and down in the analog front-end with an on-board phase locked loop at $9.05$ GHz, resulting in the lower sideband being sent to the cryostat. The analog-to-digital sampling rate is 1 GHz, and the signal is filtered by a moving average filter with a $\sim 20$~ns window. The demodulated signal is rotated onto the in-phase axis. The real-time trigger, implemented in the FPGA fabric, is activated when the imaginary quadrature of the demodulated signal surpasses a value corresponding to a phase of $15^\circ$ for all active resonators within a $10\;\upmu$s window. Upon triggering, the FPGA stores 1024 samples of the time trace of both quadratures before the trigger.

\section{Correlation length} \label{sec_corr}

Resonators in this work are operated in the bifurcated regime. This is necessary to ensure a viable signal-to-noise ratio given the fast integration time of $20$~ns (cf. Appendix \ref{baseband_stage}). An example of a reconstructed impact is shown in Fig.~\ref{fig_setup}c, and correlated phase shifts of the resonators are shown in Fig.~\ref{fig_setup}d. We fit the initial part of the response to a piecewise function composed of a baseline followed by a linear rise. We measure a quasi-diffusive propagation velocity of $3.2 \pm 1.1$ km/s (cf. Fig.~\ref{fig_setup}e).

We summarize key elements suggesting a decaying correlation length in Fig.~\ref{fig_length}. In Fig.~\ref{fig_length}a, we show a scatter plot of the reconstructed positions of the impacts measured over a $\sim 27$ hour window. The positions are reported as exact and uncertainties get encoded in the deviation of the propagation velocity. Note that the impact points all fit within the chip dimensions, despite no fitting boundaries being imposed on the $x,y$ coordinates, confirming the quality of the position reconstruction method. Furthemore, note that the impacts seem to bunch around the centroid of the resonator grid, in contrast to the expected impact flux, which should be uniform over the whole chip \cite{barnett1996review}. This is explained by the fact that the phonon wavefront disperses its energy as it travels across the chip, and with increasing distance it is increasingly likely that the energy deposited in the resonators is not sufficient to activate the trigger.

To further confirm this distance scaling, we report burst rates in Fig.~\ref{fig_length}b for different triggering protocols and environmental radioactivity conditions. We report histograms of the waiting time between bursts, showing the expected Poisson distribution, together with the exponential fit used to extract the burst rate. The rate of registered impacts increases fourfold when we relax the requirements to record a burst on triggering four neighboring resonators only. When we allow bursts to be recorded on triggering a single resonator, the measured burst rate further increases to $\sim40$~mHz (not depicted). Restricting the number of resonators that need to be triggered results in a smaller area of correlation. The observed increase in rate further seems to suggest the existence of a finite correlation length. As a sanity check, we boost the naturally occurring radioactivity of the laboratory environment by adding a cobalt-57 source in the vicinity of the cryostat. This results in an increase in correlated burst rates when triggering on four and six resonators, and with the rate on 4-resonator triggering still being the largest one. 

To assess quantitatively the correlation length, in Fig.~\ref{fig_length}c we show the slope of the phase rise as a proxy for the energy absorbed by each resonator (cf. Eq.~\eqref{eq_shift}) as a function of the distance between resonator and impact point, for all impacts reported in panel a. The data suggests that the energy of phonons reaching the resonators decays over a millimetric length, comparable to Ref.~\cite{moshe_speed}. To confirm that this is indeed the scaling visible in panel a, we show a kernel density estimation of the measured impact flux on the chip in Fig.~\ref{fig_length}d. This is obtained by summing normalized, symmetric Gaussians centered at the reconstructed impact positions (black markers in panel a) with a width that encodes the uncertainty on the propagation velocity. The Gaussian width is spanned by the two extremal distances obtained by recalculating the impact position using propagation velocities spanning a standard deviation about the mean (cf. Fig.~\ref{fig_setup}e). In the left panel we report the measured flux, showing the same spatial distribution of panel a. In the right panel we show the same flux scaled by $\exp(-d/1$ mm), where $d$ is the distance from the centroid of the resonator grid. For clarity, we only show it over the portion of the chip in which we actually measured impacts. This unbiased flux more closely resembles the expected one, with a uniform spatial distribution and a rate of a few impacts per minute per square centimeter~\cite{barnett1996review}. 

\section{Conclusions}

We reported measurements of correlated quasiparticle bursts from ionizing radiation in a superconducting quantum device made of six resonators in a multiplexed readout scheme. We reconstructed the position of the high-energy particle impacts on the chip, and inferred the propagation velocity of the generated athermal phonons. With these data at hand, we uncovered a quasiparticle burst correlation length on the order of $1$~mm for the case of a bare sapphire substrate. This suggests that, in larger (several cm$^2$) quantum processor chips, the correlated nature of most errors due to ionizing radiation is confined to a subset of qubits within a few mm$^2$. It is important to note that, for the sake of simplicity, our results were obtained on an empty substrate. Adding various metalization layers, as is the case in a quantum processor, could increase or decrease this correlation length, depending on the quasiparticle-phonon generation-recombination dynamics. Therefore, possible future research avenues include the quantification of the decay length of correlated qubit errors in actual quantum processors (similar to Refs.~\cite{Wilen2021, Acharya2023}), and measurements in the spirit of Ref.~\cite{moshe_speed} dedicated to linking the propagation velocity to the phonon energy. 

\subsection*{Data availability}
All relevant data are available from the authors upon reasonable request.

\subsection*{Competing interest}
The authors declare no competing interest.

\section*{Acknowledgements}
We acknowledge Patrick Winkel, Simon Geisert, Nicolas Zapata, Nicolas Gosling and Mitchell Field for fruitful discussion, and Lucas Radtke and Silvia Diewald for technical support. Facilities use is supported by the KIT Nanostructure Service Laboratory (NSL). We acknowledge funding from Google via their academic support initiative. F.V., P.P., T.R. and I.M.P. acknowledge support from the German Ministry of Education and Research (BMBF) within project GEQCOS (FKZ: 13N15683). A.M. acknowledges support from the Alexander von Humboldt foundation in the framework of a Research Fellowship endowed by the German Federal Ministry of Education and Research.

\section*{Appendices} \label{sec_suppl}
\appendix
\setcounter{figure}{0}
\renewcommand{\thefigure}{S\arabic{figure}}
\renewcommand{\theHfigure}{A\arabic{figure}}

\section{Custom room-temperature electronics} \label{app_rf}

The room-temperature electronics platform depicted on the right side of Fig.~\ref{fig_setup}b is a superheterodyne setup divided in two stages. First, the baseband stage where the signal is generated using a AMD Xilinx RFSoC evaluation platform \cite{zcu111} and a custom designed signal fanout board (cf. Appendix \ref{baseband_stage}). The XCZU28 RFSoC device contains eight \SI{6.5}{\giga\samples\per\second} digital-to-analog converters (DACs) and eight \SI{4}{\giga\samples\per\second} analog-to-digital converters (ADCs); both driven at \SI{4}{\giga\samples\per\second}, with the data processing bandwidth set at \SI{1}{\giga\samples\per\second}. 
Second, the Intermediate and Radio Frequency (RF) stage (cf. Appendix \ref{rf_stage}) is provided by a custom high-frequency conversion and analog signal conditioning front-end board \cite{Gartmann_2022}, providing vector signal generation and analysis. This setup can be used both for qubit characterization and quantum sensor measurement platforms \cite{Gebauer_2020, Redondo_2024}.

\subsection{Digital and baseband stage} \label{baseband_stage}

The FPGA firmware generates a frequency division multiplexed comb containing six peaks through numerically controlled oscillators (one per resonator) at different resonant frequencies. The synthesis window has a total bandwidth of 1 GHz by means of two \SI{500}{\mega\hertz} analog cables serving as an IQ interface. On the return path from the cryostat, the comb is mixed down to the baseband so that the ADCs can capture \SI{1}{\giga\samples\per\second} of digital data. A set of digital mixers shifts the frequency of the signal from each resonator to DC. The demodulated voltage time series is passed through six (one per channel) independent 56-coefficient finite impulse response low-pass filters. This moving average window gives an effective sampling time of 20 nanoseconds. The filter coefficients were generated using the Python-based \texttt{scipy.signal} library and according to the \texttt{kaiser\_beta} and \texttt{firwin} functions with the aforementioned 56 taps, nominal \SI{10}{\mega\hertz} cutoff, and \SI{60}{\decibel} of stop band rejection. A constant-phase offset calibration in the mixer rotates the demodulated signal onto the in-phase axis, such that the threshold trigger can work exclusively with the DC-like Q amplitude. Downsampling without aliasing can then be performed. Raw trigger data from a first-in-first-out array and subsequent points are then stored in 1024 sample windows, both at the net data rate and at a user-selectable subsampling rate (used to resolve the response of bursts with long tails). Stored values are then transferred to the RFSoC processor RAM and moved to the user PC via a custom software stack and the gRPC network framework \cite{Karcher_2021}.

\begin{figure}[t!]
    \includegraphics[width=1\columnwidth]{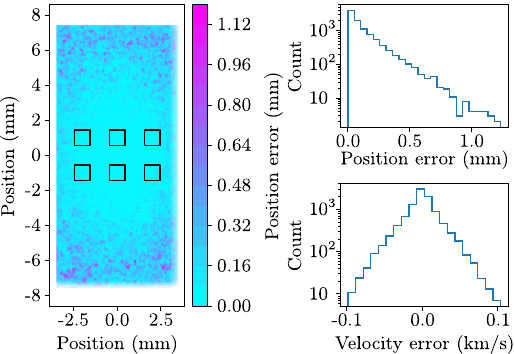}
    \caption{\textbf{Reconstruction error.} Reconstructed position and velocity errors over $10^4$ simulated impacts with a $20$ ns time discretization.} \label{fig_error}
\end{figure}

When the demodulated imaginary quadrature of the voltage surpasses a user-controllable threshold, the channel is triggered. In this work, the trigger threshold was set individually for each channel to correspond to a $15^\circ$ phase value. Upon triggering, the quadrature data (1024 nanosecond samples up to the trigger event) is saved to disk. Once triggered, the channel is locked to avoid new, potentially faulty triggers. The lock is released when fetching the data from the internal storage. The data is transferred into the analysis computer by running a Python infinite loop that constantly fetches the trigger status of the six channels, and flags as valid impact data all instances of the six channels trigger timestamps being changed from one cycle to the next. In other words, this implements an implicit cooldown time based on the average cycle time of the Python loop, which is in the millisecond range. A further check that is performed to flag data as valid is that the timestamps of all triggers for a given impact must be within a $10\;\upmu$s window to avoid triggering on longer relaxation tails (according to the spacing of the resonator grid and for a propagation velocity of 3~km/s, all bursts should happen within a $2\;\upmu$s window).

\subsection{Inter and RF stage} \label{rf_stage}

The RF signal conditioning front-end contains aliasing filters for the ADC stages of the RFSoC, mixers for the frequency translation to the desired band, local oscillators to serve as their source, some amplification for substrate loss compensation, and variable attenuation. Everything is digitally controlled through serial interfaces and interpreted by a controller daughterboard for simple user interfacing through a Python-based Ethernet control stack.

A first stage IQ mixing translates between the dual \SI{500}{\mega\hertz} baseband interface and the on-board inter-stage frequency of 2-\SI{4}{\giga\hertz}. It consists of commercial wireless transceiver circuitry providing linearity and local oscillator isolation and filtering for harmonic rejection. The inter-stage signal is connected to the RF domain through mixers with single-ended connections. The RF side signal is band-limited through a band-pass filter, the slope compensated through an equalization circuit, conversion and substrate losses compensated by amplifiers, and routed to SMA connectors.
Variable attenuators are located on the transmission side of the RF stage and the reception side of the inter-stage frequency providing signal level adjustment.

Both mixing stages receive their local oscillator signals from on-board phase-locked loop (PLL) circuits which also incorporate voltage-controlled oscillators (VCO). Sweeping can be achieved through dynamic reconfiguration of either the PLL or by adjusting the baseband input signal on the FPGA. As the reference clock is shared for both PLLs, and mixing on all stages is performed with the same VCO source, the total additive phase noise contribution to the probe tones can be kept low through the provision of an external frequency reference. 

The immediate bandwidth is given by the IQ interface, yielding \SI{1}{\giga\hertz}. The RF bandwidth spans from \SI{4}{\giga\hertz} to \SI{10}{\giga\hertz}. The dynamic range of the attenuators is \SI{31}{\decibel}, with spurious signals depending on the number of tones, desired output power, and frequency of the local oscillators.

\section{Position reconstruction} \label{app_position}

In order to determine the uncertainty of the position and velocity reconstructed from the impacts due to the finite time resolution, we simulate $10^4$ impacts uniformly distributed on the chip, each generating a phonon wavefront expanding with a velocity drawn from a normal distribution of $3 \pm 1$ km/s. The reconstructed position and velocity are well reconstructed in all cases. We simulate the effect of sampling time by recasting the continuous time variable as an integer number of $20$ ns steps | the effective sampling time given by the width of the moving average filter (cf. Appendix~\ref{baseband_stage}). In Fig.~\ref{fig_error}, we show that, even when accounting for the finite time resolution, positional errors are within  $0.1$~mm in the vicinity of the resonators (where all measured reconstructed impacts are registered), and velocity error is within $5\%$ for all simulated impacts.

\section{Resonator characterization} \label{app_reso}

 \renewcommand*{\arraystretch}{1.5}
\begin{table}[h!]
\caption{Table of resonator parameters.}
\begin{center}
  \begin{tabular}{c | c | c | c | c | c | c | c |}
     & \thead{\normalsize \shortstack{  $f_0$ \\ (GHz)      }} & 
       \thead{\normalsize \shortstack{  $Q_i$ \\ $10^5$     }} & 
       \thead{\normalsize \shortstack{  $Q_c$ \\ $10^5$     }} & 
       \thead{\normalsize \shortstack{  $\kappa / 2 \pi$ \\ (kHz)   }} & 
       \thead{\normalsize \shortstack{  $K$ \\ (mHz/photon)  }} & 
       \thead{\normalsize \shortstack{  $n_c$ \\ $10^6$     }} & 
       \thead{\normalsize \shortstack{  $\bar{n}$ \\ $10^6$ }} \\ \hline
    \cellcolor{R0}{ } & $8.708$ & $4.73$ & $0.89$ & $120$ & $-133$ & $0.514$ & $10.1$ \\
    \cellcolor{R1}{ } & $8.772$ & $5.72$ & $0.89$ & $116$ & $-135$ & $0.497$ & $10.5$ \\
    \cellcolor{R2}{ } & $8.828$ & $11.2$ & $1.47$ & $71$ & $-116$ & $0.357$ & $17.8$ \\
    \cellcolor{R3}{ } & $8.883$ & $4.44$ & $0.40$ & $242$ & $-165$ & $0.848$ & $5.2$ \\
    \cellcolor{R4}{ } & $8.943$ & $3.03$ & $0.27$ & $357$ & $-195$ & $1.05$ & $3.4$ \\
    \cellcolor{R5}{ } & $8.989$ & $2.03$ & $0.61$ & $194$ & $-131$ & $0.85$ & $5.4$ \\
    \hline
  \end{tabular}\label{table_parameters}
\end{center}
\end{table} 

We design our resonators to be spaced $50$~MHz apart, by sweeping the electrode pitch (cf. Fig.~\ref{fig_setup}a). We validate the design by performing finite elements simulations using Ansyss HFSS. We characterize the resonators in frequency domain using a Keysight Vector Network Analyzer (VNA) to perform microwave reflection measurements, and by fitting the scattering parameters using Qkit \cite{qkitcirclefit}, based on Ref.~\cite{probst2015efficient}, allowing us to extract the resonant frequency as well as the internal and coupling quality factors. Extracted parameters are reported in Table~\ref{table_parameters}. The measured resonant frequencies are spaced $56\pm 6$~MHz apart, in agreement with the designed ones. Note that the main measurement run took place 2/3 months after the VNA characterization, and as such we expect actual resonant frequencies to be on the order of $10$|$15$ MHz lower due to oxidation aging \cite{valenti_cardani_demetra}.

We characterize nonlinear behavior by sweeping the probe power, and add the attenuation on the line down (circa 84 dB, measured with the cryostat open and constituted by 60 dB of nominal attenuation plus wiring and insertion loss) to obtain the power $P_\mathrm{cold}$ at the chip input. This, together with the resonator parameters, allows us to extract the average number of circulating photons using $\bar{n} = 4 Q^2 P_\mathrm{cold} / ( Q_c \hbar \omega _0^2)$. The resonant frequency of the resonators changes as a function of the readout photons due to the Kerr shift, resulting in a $K=150 \pm 20$~mHz/photon shift across all resonators. This is in reasonable agreement with the ab initio value of $230$~mHz/photon calculated using the grAl microscopic model from Ref.~\cite{Maleeva2018}. Furthermore, this allows us to estimate the critical number of photons at which bifurcation occurs as $n_c = \kappa / (2\pi\sqrt{3}K).$ \cite{eichler2014controlling,valenti2019interplay}. The seismometry measurements are performed by operating the system in the deeply nonlinear regime (cf. rightmost two columns of Table~\ref{table_parameters}) in order to attain a viable signal-to-noise-ratio despite the short integration times. We posit that this does not affect the fact that, for a small enough time after a burst, the phase rises linearly as in the case of resonator operated in the non-bifurcated regime \cite{Swenson2010,moshe_speed}. Nevertheless, characterizing the nonlinear response to extract the value of energy deposited into the resonator is a valid research avenue for future development. Furthermore, one could employ a quantum limited amplifier to boost the signal, allowing resonators to be operated in the linear regime and Eq.~\eqref{eq_shift} to be solved directly, thus reporting the actual energy absorbed in the resonator.

\bibliography{main_refe_oneauth}

\begin{thebibliography}{35}%
\makeatletter
\providecommand \@ifxundefined [1]{%
 \@ifx{#1\undefined}
}%
\providecommand \@ifnum [1]{%
 \ifnum #1\expandafter \@firstoftwo
 \else \expandafter \@secondoftwo
 \fi
}%
\providecommand \@ifx [1]{%
 \ifx #1\expandafter \@firstoftwo
 \else \expandafter \@secondoftwo
 \fi
}%
\providecommand \natexlab [1]{#1}%
\providecommand \enquote  [1]{``#1''}%
\providecommand \bibnamefont  [1]{#1}%
\providecommand \bibfnamefont [1]{#1}%
\providecommand \citenamefont [1]{#1}%
\providecommand \href@noop [0]{\@secondoftwo}%
\providecommand \href [0]{\begingroup \@sanitize@url \@href}%
\providecommand \@href[1]{\@@startlink{#1}\@@href}%
\providecommand \@@href[1]{\endgroup#1\@@endlink}%
\providecommand \@sanitize@url [0]{\catcode `\\12\catcode `\$12\catcode `\&12\catcode `\#12\catcode `\^12\catcode `\_12\catcode `\%12\relax}%
\providecommand \@@startlink[1]{}%
\providecommand \@@endlink[0]{}%
\providecommand \url  [0]{\begingroup\@sanitize@url \@url }%
\providecommand \@url [1]{\endgroup\@href {#1}{\urlprefix }}%
\providecommand \urlprefix  [0]{URL }%
\providecommand \Eprint [0]{\href }%
\providecommand \doibase [0]{https://doi.org/}%
\providecommand \selectlanguage [0]{\@gobble}%
\providecommand \bibinfo  [0]{\@secondoftwo}%
\providecommand \bibfield  [0]{\@secondoftwo}%
\providecommand \translation [1]{[#1]}%
\providecommand \BibitemOpen [0]{}%
\providecommand \bibitemStop [0]{}%
\providecommand \bibitemNoStop [0]{.\EOS\space}%
\providecommand \EOS [0]{\spacefactor3000\relax}%
\providecommand \BibitemShut  [1]{\csname bibitem#1\endcsname}%
\let\auto@bib@innerbib\@empty
\bibitem [{\citenamefont {Thorbeck}\ \emph {et~al.}(2023)\citenamefont {Thorbeck} \emph {et~al.}}]{thorbeck_radiation}%
  \BibitemOpen
  \bibfield  {author} {\bibinfo {author} {\bibfnamefont {T.}~\bibnamefont {Thorbeck}} \emph {et~al.},\ }\bibfield  {title} {\bibinfo {title} {Two-level-system dynamics in a superconducting qubit due to background ionizing radiation},\ }\href@noop {} {\bibfield  {journal} {\bibinfo  {journal} {PRX Quantum}\ }\textbf {\bibinfo {volume} {4}},\ \bibinfo {pages} {020356} (\bibinfo {year} {2023})}\BibitemShut {NoStop}%
\bibitem [{\citenamefont {Veps{\"a}l{\"a}inen}\ \emph {et~al.}(2020)\citenamefont {Veps{\"a}l{\"a}inen} \emph {et~al.}}]{Vepslinen2020impact}%
  \BibitemOpen
  \bibfield  {author} {\bibinfo {author} {\bibfnamefont {A.~P.}\ \bibnamefont {Veps{\"a}l{\"a}inen}} \emph {et~al.},\ }\bibfield  {title} {\bibinfo {title} {Impact of ionizing radiation on superconducting qubit coherence},\ }\href@noop {} {\bibfield  {journal} {\bibinfo  {journal} {Nature}\ }\textbf {\bibinfo {volume} {584}},\ \bibinfo {pages} {551} (\bibinfo {year} {2020})}\BibitemShut {NoStop}%
\bibitem [{\citenamefont {Cardani}\ \emph {et~al.}(2021)\citenamefont {Cardani}, \citenamefont {Valenti} \emph {et~al.}}]{valenti_cardani_demetra}%
  \BibitemOpen
  \bibfield  {author} {\bibinfo {author} {\bibfnamefont {L.}~\bibnamefont {Cardani}}, \bibinfo {author} {\bibfnamefont {F.}~\bibnamefont {Valenti}}, \emph {et~al.},\ }\bibfield  {title} {\bibinfo {title} {Reducing the impact of radioactivity on quantum circuits in a deep-underground facility},\ }\href@noop {} {\bibfield  {journal} {\bibinfo  {journal} {Nature Communications}\ }\textbf {\bibinfo {volume} {12}},\ \bibinfo {pages} {2733} (\bibinfo {year} {2021})}\BibitemShut {NoStop}%
\bibitem [{\citenamefont {Larson}\ \emph {et~al.}(2025)\citenamefont {Larson} \emph {et~al.}}]{larson2025}%
  \BibitemOpen
  \bibfield  {author} {\bibinfo {author} {\bibfnamefont {C.~P.}\ \bibnamefont {Larson}} \emph {et~al.},\ }\bibfield  {title} {\bibinfo {title} {Quasiparticle poisoning of superconducting qubits with active gamma irradiation},\ }\href@noop {} {\bibfield  {journal} {\bibinfo  {journal} {arXiv:2503.07354}\ } (\bibinfo {year} {2025})}\BibitemShut {NoStop}%
\bibitem [{\citenamefont {Swenson}\ \emph {et~al.}(2010)\citenamefont {Swenson} \emph {et~al.}}]{Swenson2010}%
  \BibitemOpen
  \bibfield  {author} {\bibinfo {author} {\bibfnamefont {L.~J.}\ \bibnamefont {Swenson}} \emph {et~al.},\ }\bibfield  {title} {\bibinfo {title} {High-speed phonon imaging using frequency-multiplexed kinetic inductance detectors},\ }\href@noop {} {\bibfield  {journal} {\bibinfo  {journal} {Appl. Phys. Lett.}\ }\textbf {\bibinfo {volume} {96}},\ \bibinfo {pages} {263511} (\bibinfo {year} {2010})}\BibitemShut {NoStop}%
\bibitem [{\citenamefont {Karatsu}\ \emph {et~al.}(2019)\citenamefont {Karatsu} \emph {et~al.}}]{Karatsu2019}%
  \BibitemOpen
  \bibfield  {author} {\bibinfo {author} {\bibfnamefont {K.}~\bibnamefont {Karatsu}} \emph {et~al.},\ }\bibfield  {title} {\bibinfo {title} {Mitigation of cosmic ray effect on microwave kinetic inductance detector arrays},\ }\href@noop {} {\bibfield  {journal} {\bibinfo  {journal} {Appl. Phys. Lett.}\ }\textbf {\bibinfo {volume} {114}},\ \bibinfo {pages} {032601} (\bibinfo {year} {2019})}\BibitemShut {NoStop}%
\bibitem [{\citenamefont {Wilen}\ \emph {et~al.}(2021)\citenamefont {Wilen}, \citenamefont {Abdullah} \emph {et~al.}}]{Wilen2021}%
  \BibitemOpen
  \bibfield  {author} {\bibinfo {author} {\bibfnamefont {C.~D.}\ \bibnamefont {Wilen}}, \bibinfo {author} {\bibfnamefont {S.}~\bibnamefont {Abdullah}}, \emph {et~al.},\ }\bibfield  {title} {\bibinfo {title} {Correlated charge noise and relaxation errors in superconducting qubits},\ }\href@noop {} {\bibfield  {journal} {\bibinfo  {journal} {Nature}\ }\textbf {\bibinfo {volume} {594}},\ \bibinfo {pages} {369} (\bibinfo {year} {2021})}\BibitemShut {NoStop}%
\bibitem [{\citenamefont {McEwen}\ \emph {et~al.}(2022)\citenamefont {McEwen} \emph {et~al.}}]{McEwen2022}%
  \BibitemOpen
  \bibfield  {author} {\bibinfo {author} {\bibfnamefont {M.}~\bibnamefont {McEwen}} \emph {et~al.},\ }\bibfield  {title} {\bibinfo {title} {Resolving catastrophic error bursts from cosmic rays in large arrays of superconducting qubits},\ }\href@noop {} {\bibfield  {journal} {\bibinfo  {journal} {Nature Physics}\ }\textbf {\bibinfo {volume} {18}},\ \bibinfo {pages} {107} (\bibinfo {year} {2022})}\BibitemShut {NoStop}%
\bibitem [{\citenamefont {Yelton}\ \emph {et~al.}(2024)\citenamefont {Yelton} \emph {et~al.}}]{yelton_modeling}%
  \BibitemOpen
  \bibfield  {author} {\bibinfo {author} {\bibfnamefont {E.}~\bibnamefont {Yelton}} \emph {et~al.},\ }\bibfield  {title} {\bibinfo {title} {Modeling phonon-mediated quasiparticle poisoning in superconducting qubit arrays},\ }\href@noop {} {\bibfield  {journal} {\bibinfo  {journal} {Phys. Rev. B}\ }\textbf {\bibinfo {volume} {110}},\ \bibinfo {pages} {024519} (\bibinfo {year} {2024})}\BibitemShut {NoStop}%
\bibitem [{\citenamefont {Henriques}\ \emph {et~al.}(2019)\citenamefont {Henriques}, \citenamefont {Valenti} \emph {et~al.}}]{HenriquesValenti2019}%
  \BibitemOpen
  \bibfield  {author} {\bibinfo {author} {\bibfnamefont {F.}~\bibnamefont {Henriques}}, \bibinfo {author} {\bibfnamefont {F.}~\bibnamefont {Valenti}}, \emph {et~al.},\ }\bibfield  {title} {\bibinfo {title} {Phonon traps reduce the quasiparticle density in superconducting circuits},\ }\href@noop {} {\bibfield  {journal} {\bibinfo  {journal} {Appl. Phys. Lett.}\ }\textbf {\bibinfo {volume} {115}},\ \bibinfo {pages} {212601} (\bibinfo {year} {2019})}\BibitemShut {NoStop}%
\bibitem [{\citenamefont {Martinis}(2020)}]{martinis2020}%
  \BibitemOpen
  \bibfield  {author} {\bibinfo {author} {\bibfnamefont {J.}~\bibnamefont {Martinis}},\ }\bibfield  {title} {\bibinfo {title} {Saving superconducting quantum processors from qubit decay and correlated errors generated by gamma and cosmic rays},\ }\href@noop {} {\bibfield  {journal} {\bibinfo  {journal} {arXiv:2012.06137}\ } (\bibinfo {year} {2020})}\BibitemShut {NoStop}%
\bibitem [{\citenamefont {McEwen}\ \emph {et~al.}(2024)\citenamefont {McEwen}, \citenamefont {Miao} \emph {et~al.}}]{McEwen2024Feb}%
  \BibitemOpen
  \bibfield  {author} {\bibinfo {author} {\bibfnamefont {M.}~\bibnamefont {McEwen}}, \bibinfo {author} {\bibfnamefont {K.~C.}\ \bibnamefont {Miao}}, \emph {et~al.},\ }\bibfield  {title} {\bibinfo {title} {Resisting high-energy impact events through gap engineering in superconducting qubit arrays},\ }\href@noop {} {\bibfield  {journal} {\bibinfo  {journal} {Phys. Rev. Lett.}\ }\textbf {\bibinfo {volume} {133}},\ \bibinfo {pages} {240601} (\bibinfo {year} {2024})}\BibitemShut {NoStop}%
\bibitem [{\citenamefont {Acharya}\ \emph {et~al.}(2025)\citenamefont {Acharya} \emph {et~al.}}]{Acharya2024Aug}%
  \BibitemOpen
  \bibfield  {author} {\bibinfo {author} {\bibfnamefont {R.}~\bibnamefont {Acharya}} \emph {et~al.},\ }\bibfield  {title} {\bibinfo {title} {Quantum error correction below the surface code threshold},\ }\href@noop {} {\bibfield  {journal} {\bibinfo  {journal} {Nature}\ }\textbf {\bibinfo {volume} {638}},\ \bibinfo {pages} {920} (\bibinfo {year} {2025})}\BibitemShut {NoStop}%
\bibitem [{\citenamefont {Moshel}\ \emph {et~al.}(2024)\citenamefont {Moshel}, \citenamefont {Rabinowitz}, \citenamefont {Blumenthal},\ and\ \citenamefont {Hacohen-Gourgy}}]{moshe_speed}%
  \BibitemOpen
  \bibfield  {author} {\bibinfo {author} {\bibfnamefont {G.}~\bibnamefont {Moshel}}, \bibinfo {author} {\bibfnamefont {O.}~\bibnamefont {Rabinowitz}}, \bibinfo {author} {\bibfnamefont {E.}~\bibnamefont {Blumenthal}},\ and\ \bibinfo {author} {\bibfnamefont {S.}~\bibnamefont {Hacohen-Gourgy}},\ }\bibfield  {title} {\bibinfo {title} {{Propagation velocity measurements of substrate phonon bursts using MKIDs for superconducting circuits}},\ }\href@noop {} {\bibfield  {journal} {\bibinfo  {journal} {Applied Physics Letters}\ }\textbf {\bibinfo {volume} {125}},\ \bibinfo {pages} {222601} (\bibinfo {year} {2024})}\BibitemShut {NoStop}%
\bibitem [{\citenamefont {Winkel}\ \emph {et~al.}(2020)\citenamefont {Winkel} \emph {et~al.}}]{winkel_transmon}%
  \BibitemOpen
  \bibfield  {author} {\bibinfo {author} {\bibfnamefont {P.}~\bibnamefont {Winkel}} \emph {et~al.},\ }\bibfield  {title} {\bibinfo {title} {Implementation of a transmon qubit using superconducting granular aluminum},\ }\href@noop {} {\bibfield  {journal} {\bibinfo  {journal} {Phys. Rev. X}\ }\textbf {\bibinfo {volume} {10}},\ \bibinfo {pages} {031032} (\bibinfo {year} {2020})}\BibitemShut {NoStop}%
\bibitem [{\citenamefont {Day}\ \emph {et~al.}(2003)\citenamefont {Day} \emph {et~al.}}]{day2003broadband}%
  \BibitemOpen
  \bibfield  {author} {\bibinfo {author} {\bibfnamefont {P.~K.}\ \bibnamefont {Day}} \emph {et~al.},\ }\bibfield  {title} {\bibinfo {title} {A broadband superconducting detector suitable for use in large arrays},\ }\href@noop {} {\bibfield  {journal} {\bibinfo  {journal} {Nature}\ }\textbf {\bibinfo {volume} {425}},\ \bibinfo {pages} {817} (\bibinfo {year} {2003})}\BibitemShut {NoStop}%
\bibitem [{\citenamefont {Cardani}\ \emph {et~al.}(2023)\citenamefont {Cardani} \emph {et~al.}}]{Cardani2023Jan}%
  \BibitemOpen
  \bibfield  {author} {\bibinfo {author} {\bibfnamefont {L.}~\bibnamefont {Cardani}} \emph {et~al.},\ }\bibfield  {title} {\bibinfo {title} {{Disentangling the sources of ionizing radiation in superconducting qubits}},\ }\href@noop {} {\bibfield  {journal} {\bibinfo  {journal} {Eur. Phys. J. C}\ }\textbf {\bibinfo {volume} {83}},\ \bibinfo {pages} {1} (\bibinfo {year} {2023})}\BibitemShut {NoStop}%
\bibitem [{\citenamefont {Cheng}\ \emph {et~al.}(2020)\citenamefont {Cheng} \emph {et~al.}}]{sapphiredos}%
  \BibitemOpen
  \bibfield  {author} {\bibinfo {author} {\bibfnamefont {Z.}~\bibnamefont {Cheng}} \emph {et~al.},\ }\bibfield  {title} {\bibinfo {title} {Thermal conductance across harmonic-matched epitaxial al-sapphire heterointerfaces},\ }\href@noop {} {\bibfield  {journal} {\bibinfo  {journal} {Communications Physics}\ }\textbf {\bibinfo {volume} {3}} (\bibinfo {year} {2020})}\BibitemShut {NoStop}%
\bibitem [{\citenamefont {Ogburn}(2008)}]{ogburn2008thesis}%
  \BibitemOpen
  \bibfield  {author} {\bibinfo {author} {\bibfnamefont {R.~W.}\ \bibnamefont {Ogburn}},\ }\emph {\bibinfo {title} {\textup{A search for particle dark matter using cryogenic germanium and silicon detectors in the one- and two-tower runs of cdms-ii at soudan}}},\ \href@noop {} {Ph.D. thesis},\ \bibinfo  {school} {Stanford University} (\bibinfo {year} {2008})\BibitemShut {NoStop}%
\bibitem [{sci()}]{scipylsq}%
  \BibitemOpen
  \href@noop {} {\bibinfo {title} {{Scipy.optimize library}}},\ \bibinfo {note} {{https://docs.scipy.org/doc/scipy/ reference/optimize.html}}\BibitemShut {NoStop}%
\bibitem [{\citenamefont {Kaplan}\ \emph {et~al.}(1976)\citenamefont {Kaplan} \emph {et~al.}}]{kaplan76}%
  \BibitemOpen
  \bibfield  {author} {\bibinfo {author} {\bibfnamefont {S.~B.}\ \bibnamefont {Kaplan}} \emph {et~al.},\ }\bibfield  {title} {\bibinfo {title} {Quasiparticle and phonon lifetimes in superconductors},\ }\href@noop {} {\bibfield  {journal} {\bibinfo  {journal} {Phys. Rev. B}\ }\textbf {\bibinfo {volume} {14}},\ \bibinfo {pages} {4854} (\bibinfo {year} {1976})}\BibitemShut {NoStop}%
\bibitem [{\citenamefont {Gr\"unhaupt}\ \emph {et~al.}(2018)\citenamefont {Gr\"unhaupt} \emph {et~al.}}]{Grunhaupt2018QPgral}%
  \BibitemOpen
  \bibfield  {author} {\bibinfo {author} {\bibfnamefont {L.}~\bibnamefont {Gr\"unhaupt}} \emph {et~al.},\ }\bibfield  {title} {\bibinfo {title} {Loss mechanisms and quasiparticle dynamics in superconducting microwave resonators made of thin-film granular aluminum},\ }\href@noop {} {\bibfield  {journal} {\bibinfo  {journal} {Phys. Rev. Lett.}\ }\textbf {\bibinfo {volume} {121}},\ \bibinfo {pages} {117001} (\bibinfo {year} {2018})}\BibitemShut {NoStop}%
\bibitem [{\citenamefont {Rotzinger}\ \emph {et~al.}(2016)\citenamefont {Rotzinger} \emph {et~al.}}]{Rotzinger_2016}%
  \BibitemOpen
  \bibfield  {author} {\bibinfo {author} {\bibfnamefont {H.}~\bibnamefont {Rotzinger}} \emph {et~al.},\ }\bibfield  {title} {\bibinfo {title} {Aluminium-oxide wires for superconducting high kinetic inductance circuits},\ }\href@noop {} {\bibfield  {journal} {\bibinfo  {journal} {Superconductor Science and Technology}\ }\textbf {\bibinfo {volume} {30}},\ \bibinfo {pages} {025002} (\bibinfo {year} {2016})}\BibitemShut {NoStop}%
\bibitem [{\citenamefont {Valenti}\ \emph {et~al.}(2019)\citenamefont {Valenti} \emph {et~al.}}]{valenti2019interplay}%
  \BibitemOpen
  \bibfield  {author} {\bibinfo {author} {\bibfnamefont {F.}~\bibnamefont {Valenti}} \emph {et~al.},\ }\bibfield  {title} {\bibinfo {title} {Interplay between kinetic inductance, nonlinearity, and quasiparticle dynamics in granular aluminum microwave kinetic inductance detectors},\ }\href@noop {} {\bibfield  {journal} {\bibinfo  {journal} {Phys. Rev. Applied}\ }\textbf {\bibinfo {volume} {11}},\ \bibinfo {pages} {054087} (\bibinfo {year} {2019})}\BibitemShut {NoStop}%
\bibitem [{\citenamefont {Gartmann}\ \emph {et~al.}(2022)\citenamefont {Gartmann} \emph {et~al.}}]{Gartmann_2022}%
  \BibitemOpen
  \bibfield  {author} {\bibinfo {author} {\bibfnamefont {R.}~\bibnamefont {Gartmann}} \emph {et~al.},\ }\bibfield  {title} {\bibinfo {title} {Progress of the {ECHo} {SDR} readout hardware for multiplexed {MMCs}},\ }\href@noop {} {\bibfield  {journal} {\bibinfo  {journal} {J. Low Temp. Phys.}\ } (\bibinfo {year} {2022})}\BibitemShut {NoStop}%
\bibitem [{\citenamefont {Barnett}\ \emph {et~al.}(1996)\citenamefont {Barnett} \emph {et~al.}}]{barnett1996review}%
  \BibitemOpen
  \bibfield  {author} {\bibinfo {author} {\bibfnamefont {R.~M.}\ \bibnamefont {Barnett}} \emph {et~al.},\ }\bibfield  {title} {\bibinfo {title} {Review of particle physics},\ }\href@noop {} {\bibfield  {journal} {\bibinfo  {journal} {Phys. Rev. D}\ }\textbf {\bibinfo {volume} {54}},\ \bibinfo {pages} {1} (\bibinfo {year} {1996})}\BibitemShut {NoStop}%
\bibitem [{\citenamefont {Acharya}\ \emph {et~al.}(2023)\citenamefont {Acharya} \emph {et~al.}}]{Acharya2023}%
  \BibitemOpen
  \bibfield  {author} {\bibinfo {author} {\bibfnamefont {R.}~\bibnamefont {Acharya}} \emph {et~al.},\ }\bibfield  {title} {\bibinfo {title} {Suppressing quantum errors by scaling a surface code logical qubit},\ }\href@noop {} {\bibfield  {journal} {\bibinfo  {journal} {Nature}\ }\textbf {\bibinfo {volume} {614}},\ \bibinfo {pages} {676} (\bibinfo {year} {2023})}\BibitemShut {NoStop}%
\bibitem [{zcu()}]{zcu111}%
  \BibitemOpen
  \href@noop {} {\bibinfo {title} {{ Zynq UltraScale+ RFSoC ZCU111 Evaluation Kit}}},\ \bibinfo {note} {{ https://www.xilinx.com/products/boards-and-kits/zcu111.html }}\BibitemShut {NoStop}%
\bibitem [{\citenamefont {Gebauer}\ \emph {et~al.}(2020)\citenamefont {Gebauer} \emph {et~al.}}]{Gebauer_2020}%
  \BibitemOpen
  \bibfield  {author} {\bibinfo {author} {\bibfnamefont {R.}~\bibnamefont {Gebauer}} \emph {et~al.},\ }\bibfield  {title} {\bibinfo {title} {State preparation of a fluxonium qubit with feedback from a custom {FPGA-based} platform},\ }in\ \href@noop {} {\emph {\bibinfo {booktitle} {Fifth International Conference on Quantum Technologies ({ICQT-2019})}}}\ (\bibinfo  {publisher} {AIP Publishing},\ \bibinfo {year} {2020})\BibitemShut {NoStop}%
\bibitem [{\citenamefont {García~Redondo}\ \emph {et~al.}(2024)\citenamefont {García~Redondo} \emph {et~al.}}]{Redondo_2024}%
  \BibitemOpen
  \bibfield  {author} {\bibinfo {author} {\bibfnamefont {M.~E.}\ \bibnamefont {García~Redondo}} \emph {et~al.},\ }\bibfield  {title} {\bibinfo {title} {{RFSoC} gen3-based software-defined radio characterization for the readout system of low-temperature bolometers},\ }\href@noop {} {\bibfield  {journal} {\bibinfo  {journal} {J. Low Temp. Phys.}\ }\textbf {\bibinfo {volume} {215}},\ \bibinfo {pages} {161} (\bibinfo {year} {2024})}\BibitemShut {NoStop}%
\bibitem [{\citenamefont {Karcher}\ \emph {et~al.}(2021)\citenamefont {Karcher} \emph {et~al.}}]{Karcher_2021}%
  \BibitemOpen
  \bibfield  {author} {\bibinfo {author} {\bibfnamefont {N.}~\bibnamefont {Karcher}} \emph {et~al.},\ }\bibfield  {title} {\bibinfo {title} {Versatile configuration and control framework for real-time data acquisition systems},\ }\href@noop {} {\bibfield  {journal} {\bibinfo  {journal} {IEEE Transactions on Nuclear Science}\ }\textbf {\bibinfo {volume} {68}},\ \bibinfo {pages} {1899} (\bibinfo {year} {2021})}\BibitemShut {NoStop}%
\bibitem [{qki()}]{qkitcirclefit}%
  \BibitemOpen
  \href@noop {} {\bibinfo {title} {{Qkit: a quantum measurement suite in Python}}},\ \bibinfo {howpublished} {https://github.com/qkitgroup/qkit}\BibitemShut {NoStop}%
\bibitem [{\citenamefont {Probst}\ \emph {et~al.}(2015)\citenamefont {Probst} \emph {et~al.}}]{probst2015efficient}%
  \BibitemOpen
  \bibfield  {author} {\bibinfo {author} {\bibfnamefont {S.}~\bibnamefont {Probst}} \emph {et~al.},\ }\bibfield  {title} {\bibinfo {title} {Efficient and robust analysis of complex scattering data under noise in microwave resonators},\ }\href@noop {} {\bibfield  {journal} {\bibinfo  {journal} {Review of Scientific Instruments}\ }\textbf {\bibinfo {volume} {86}},\ \bibinfo {pages} {024706} (\bibinfo {year} {2015})}\BibitemShut {NoStop}%
\bibitem [{\citenamefont {Maleeva}\ \emph {et~al.}(2018)\citenamefont {Maleeva} \emph {et~al.}}]{Maleeva2018}%
  \BibitemOpen
  \bibfield  {author} {\bibinfo {author} {\bibfnamefont {N.}~\bibnamefont {Maleeva}} \emph {et~al.},\ }\bibfield  {title} {\bibinfo {title} {Circuit quantum electrodynamics of granular aluminum resonators},\ }\href@noop {} {\bibfield  {journal} {\bibinfo  {journal} {Nature Communications}\ }\textbf {\bibinfo {volume} {9}},\ \bibinfo {pages} {3889} (\bibinfo {year} {2018})}\BibitemShut {NoStop}%
\bibitem [{\citenamefont {Eichler}\ and\ \citenamefont {Wallraff}(2014)}]{eichler2014controlling}%
  \BibitemOpen
  \bibfield  {author} {\bibinfo {author} {\bibfnamefont {C.}~\bibnamefont {Eichler}}\ and\ \bibinfo {author} {\bibfnamefont {A.}~\bibnamefont {Wallraff}},\ }\bibfield  {title} {\bibinfo {title} {{Controlling the dynamic range of a Josephson parametric amplifier}},\ }\href@noop {} {\bibfield  {journal} {\bibinfo  {journal} {EPJ Quantum Technology}\ }\textbf {\bibinfo {volume} {1}},\ \bibinfo {pages} {2} (\bibinfo {year} {2014})}\BibitemShut {NoStop}%
\end{thebibliography}%

\end{document}